\documentclass[10pt,conference]{IEEEtran}\IEEEoverridecommandlockouts
\usepackage{epsfig,graphics,subfigure,psfrag,amsmath,cases}
\usepackage{latexsym,amssymb,amsmath,epsfig,subfigure,algorithm}
\usepackage{algorithmic}
\usepackage{color}
\usepackage{url}

\author{\authorblockN{Derrick Wing Kwan Ng and Robert Schober
\thanks{This work was supported in part by the AvH Professorship
Program of the Alexander von Humboldt Foundation.}}
Institute for Digital Communications\\
 Universit\"at Erlangen-N\"urnberg, Cauerstra\ss e 7, 91058
Erlangen, Germany\\
 Email:\,\{kwan,\,schober\}@lnt.de
}

\newcommand{\abs}[1]{\lvert#1\rvert}
\newcommand{\norm}[1]{\lVert#1\rVert}

\newtheorem{Cor}{Corollary}

\newtheorem{proposition}{Proposition}
\title{Spectral Efficiency in Large-Scale MIMO-OFDM Systems with
Per-Antenna
 Power Cost}

\begin{document}
%
\maketitle
\begin{abstract}
In this paper, resource allocation for multiple-input multiple-output
orthogonal frequency division multiplexing
(MIMO-OFDM) downlink networks with large numbers of base station
antennas is studied. Assuming perfect channel state information at
the transmitter, the resource allocation algorithm design is
modeled as a non-convex optimization problem which takes into account the
joint power consumption of the power amplifiers, antenna unit, and
signal processing circuit unit. Subsequently, by exploiting
the law of large numbers and dual decomposition, an efficient suboptimal
iterative
resource allocation algorithm is proposed for maximization of the
system capacity (bit/s). In
particular, closed-form power allocation and antenna
allocation policies are derived in each iteration. Simulation
results illustrate that the proposed iterative resource allocation
algorithm achieves a close-to-optimal performance in a small number of
iterations and unveil
 a trade-off between system capacity and the number of activated
antennas: Activating all antennas may not be a good solution for
system capacity maximization when a system with a per antenna power
cost is considered.
\end{abstract}

\section{Introduction}
\label{sec:intro} The demand for ubiquitous service coverage and high
speed communications has been growing rapidly over the
last decade. Multiple-input multiple-output orthogonal frequency
division multiplexing (MIMO-OFDM) technology is considered
  as a viable solution for addressing this issue, as it provides
  extra degrees of freedom for resource allocation. Specifically, the
ergodic capacity of
a point-to-point MIMO fading channel increases practically linearly with
the
minimum of the number of transmitter and receiver antennas
\cite{book:david_wirelss_com,book:Goldsmith}.
Theoretically, the system performance can be unlimited if the
number of antennas at both the transmitter and the receiver is increased.
However, the
complexity of MIMO receivers limits the performance gains that can be
achieved in practical systems, especially for small handheld devices. In
order to facilitate the MIMO technology in practice, an
alternative form of MIMO has been proposed where a base station (BS)
with a large number of antennas serves one single antenna user
\cite{JR:large_number_antennas}-\nocite{JR:large_antennas-constant-en,JR:LA-detection,JR:LA4,CN:LA_measure1}\cite{CN:LA_measure2}.
 This new paradigm for multiple antenna systems is known as massive MIMO
 which does not only shift the signal processing burden from the
receivers to the BS, but also provides a promising system performance.
As a result, massive MIMO system design has recently
drawn much attention from both industry and academia
\cite{JR:large_number_antennas}-\cite{CN:LA_measure2}.

 In \cite{JR:large_number_antennas}, high throughput gains were shown to
be achievable in both
uplink and downlink for a time division duplex (TDD) system with massive
MIMO.
In \cite{JR:large_antennas-constant-en}, the authors derived an
achievable downlink data rate for a single-user
massive MIMO channel under the constraint of a constant signal envelope
per antenna.
 On the other hand, different receiver signal detection algorithms were
studied and channel measurements were carried out for facilitating the
implementation of massive MIMO. In \cite{JR:LA-detection} and
\cite{JR:LA4}, low complexity likelihood ascent search algorithms were
proposed for signal detection in large-scale MIMO multicarrier and single
carrier systems, receptively. In \cite{CN:LA_measure1} and
\cite{CN:LA_measure2}, channel measurements for massive MIMO systems
were reported which revealed the actual potential system performance of
large-scale antenna systems. A substantial capacity gain
(bit/s/Hz) was observed with
massive MIMO compared to single antenna systems in all studies
\cite{JR:large_number_antennas}-\cite{CN:LA_measure2}. Yet, the
advantages of massive MIMO do not come for free and the number of
antennas cannot be unlimited. In practice, each extra antenna
imposes an additional power cost on the power budget due to
the associated electronic circuitries
\cite{CN:power_consumption_elements}. Besides, both the power
amplifiers (PAs) and the antenna circuits are sharing the same finite
capacity power source at the BS. This joint power consumption model
has not been taken into account in the literature so far. Therefore, the
results in
\cite{JR:large_number_antennas}-\cite{CN:LA_measure2} which are valid
for unlimited numbers of antennas and unlimited power supply, may no
longer be applicable for the case of finite numbers of antennas and
limited power supply.

 In this paper, we address the above issues. In Section \ref{sect:OFDM},
we introduce the adopted MIMO-OFDM channel. In Section \ref{sect:forumlation}, we
formulate the resource allocation algorithm design as an
optimization problem and maximize the system capacity for
communication in MIMO-OFDM systems. To this end, an suboptimal iterative
algorithm is proposed in Section \ref{sect:solution}. In Section \ref{sect:5}, we show that the proposed suboptimal
algorithm does not only have a fast convergence, but also achieves a
close-to-optimal performance. In
Section \ref{sect:conclusion}, we conclude with a brief summary of
our results.

\section{MIMO-OFDM System Model}\label{sect:OFDM}

In this section, after introducing the notation used in this paper,
we present the adopted system model.
\subsection{Notation}
 A complex Gaussian random variable with mean $\mu$ and variance
$\sigma^2$ is denoted by ${\cal CN}(\mu,\sigma^2)$.
$\big[x\big]^+=\max\{0,x\}$. $\big[x\big]^a_b=a,\ \mbox{if}\
x>a,\big[x\big]^a_b=x,\mbox{ if}\ b\le x\le a,\big[x\big]^a_b
=b,\ \mbox{if}\ b>x$. $\big[x\big]^+=\max\{0,x\}$.
$\abs{\cdot}$ and $\norm{\cdot}$ denote the absolute
value of a complex-valued scalar and the Euclidean norm of a vector,
respectively. $\mathbb{C}^{N\times M}$ is the space of all $N\times
M$ matrices
with complex entries and $[\cdot]^T$ denotes the transpose
operation.
\subsection{MIMO-OFDM Downlink Channel Model} \label{sect:channel_model}
\begin{figure}[t]
\includegraphics[width=3.5in]{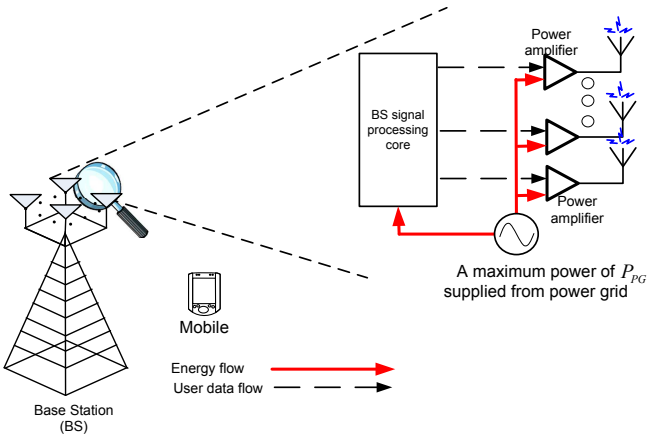}
\caption{Illustration of a MIMO-OFDM downlink network and power supply
model. }\label{fig:BS}
\end{figure}

 We consider a MIMO-OFDM
system which consists of a BS with multiple antennas and a mobile
user equipped with a single antenna, cf. Figure \ref{fig:BS}. The total
 bandwidth of the system is $\cal B$ Hertz and there are $n_F$
subcarriers. Each subcarrier has  a bandwidth $W={\cal B}/n_F$ Hertz. We assume a TDD system and the
downlink channel gains can be accurately obtained by measuring the
uplink channel based on channel reciprocity. The channel impulse
response is assumed to be time invariant (slow fading). The downlink
received symbol at the user
on subcarrier $i\in\{1,\,\ldots,\,n_F\}$ is
given by
\begin{eqnarray}
y_{i}=\sqrt{P_{i}l g}\mathbf{h}_{i}^T\mathbf{
f}_{i}x_{i}+z_{i},
\end{eqnarray}
where $x_{i}$, $P_{i}$, and $\mathbf{
f}_{i}\in\mathbb{C}^{N_{T_{i}}\times 1}$ are the transmitted
symbol, transmitted power, and precoding vector for the link from
the BS to the user on subcarrier $i$, respectively.
 $N_{T_{i}}$ is the number of active antennas allocated
on subcarrier $i$ for transmission.
$\mathbf{h}_{i}\in\mathbb{C}^{N_{T_{i}} \times 1}$ is the vector
of multipath fading coefficients between the BS and the user. The
elements in $\mathbf{h}_{i}$ are assumed to be independent and
identically distributed (i.i.d.). $l$ and $g$ represent the
path loss and shadowing between the BS and the user, respectively.
 $z_{i}$ is additive white Gaussian noise (AWGN) on subcarrier $i$ with
${\cal CN}(0,N_0W)$, where $N_0$ is  the  power spectral density.

\section{Resource Allocation Algorithm Design}\label{sect:forumlation}
In this section, we introduce the adopted system performance metric
and formulate the corresponding resource allocation  problem.
\subsection{Instantaneous Channel Capacity}
\label{subsect:Instaneous_Mutual_information}
In this subsection, we define the adopted system performance
measure. Given perfect channel state information (CSI) at the
receiver, the channel capacity between the BS and the user on
subcarrier $i$ with channel bandwidth $W={\cal B}/n_F$ is given by
   \begin{eqnarray}\label{eqn:cap}
C_{i}=W\log_2\Big(1+\Gamma_{i}\Big)\,\,
\mbox{and}\,\,
\Gamma_{i}\hspace*{-0.5mm}=\hspace*{-0.5mm}\frac{P_{i}l
g\abs{\mathbf{h}_{i}^T
\mathbf{f}_{i}}^2}{N_0W},
\end{eqnarray}
where $\Gamma_{i}$ is the received signal-to-noise ratio (SNR) at
the user on subcarrier $i$.

The \emph{aggregate system capacity} is defined as a weighted sum
of the number of bits per second successfully delivered to the
mobile users (bits-per-second) and is given by
\begin{eqnarray}
 \label{eqn:avg-sys-goodput} && U({\cal P},{\cal A})=\sum_{i=1}^{n_F}
C_{i},
\end{eqnarray}
where ${\cal P}$ and $\cal A$ are the power allocation policy and the
antenna allocation policy,
respectively.

\subsection{Power Consumption Model}
We model the power
dissipation in the system as the sum of three terms
\cite{CN:energy_efficient_2} which can be expressed as
\begin{eqnarray}
 \label{eqn:power_consumption} U_{TP}({\cal P},{\cal A})
&=&U({\cal A})+U({\cal P})+P_0\quad \mbox{where}\\
U({\cal
A})&=&\hspace*{-22mm}\underbrace{\max_{i}\,\{N_{T_{i}}\}\times
P_{A_C}}_{\mbox{Circuit power consumption of all antennas at the
BS}}\, \mbox{and}\notag\\
U({\cal P})&=&\hspace*{-22mm}\underbrace{\sum_{i=1}^{n_F}\varepsilon
P_{i}}_{\mbox{Power consumption of all power amplifiers at the BS}}.
\end{eqnarray}
 $P_{A_C}$ is the constant \emph{circuit power consumption per
antenna}, which includes the power dissipation of the transmit
filter, mixer, frequency synthesizer, and digital-to-analog
converter and is independent of the actual transmitted power. In the
considered system, we assume that there is a maximum and a minimum number of active antennas, i.e.,
 $N_{\max}$ and $N_{\min}$, at the BS. However, we do not necessarily
activate the maximum number of antennas for the sake of efficient
communication and the optimal
number of active antennas will be found in the next section based on
optimization. The physical meaning of the term
$\underset{i}{\max}\,\{ N_{T_{i}}\}$ in
(\ref{eqn:power_consumption}) is that if an antenna is activated, it
consumes power even if it is used only by some
of the subcarriers. In other words, the first term in
(\ref{eqn:power_consumption}) represents the total power consumed by
the activated antennas. The second term in
(\ref{eqn:power_consumption}) denotes the total power consumption in
the radio frequency (RF) PAs of the BS.
$\varepsilon\ge
1$ is a constant which accounts for the inefficiency in the power
amplifier and the power efficiency is defined as
$\frac{1}{\varepsilon}$. $P_0$ is the basic signal processing power
consumption which is
independent of the number of transmit antennas.
\subsection{Optimization Problem Formulation}
\label{sect:cross-Layer_formulation}
The optimal resource allocation policies (${\cal P}^*,{\cal
A}^*$) can be obtained by solving
\begin{eqnarray}
\label{eqn:cross-layer}&& \qquad \max_{{\cal P},{\cal A}
}\,\, U({\cal P},{\cal A}) \\
\notag && \hspace*{-6mm}\mbox{s.t.}\,\, \mbox{C1:}
\sum_{i=1}^{n_F}P_{i}\le P_{{\max}},\\
\notag &&\hspace*{-1mm}\mbox{C2: }
U_{TP}({\cal P},{\cal A})\le P_{PG},\hspace*{2mm} \mbox{C3:
}P_{i}\ge 0,\, \forall i,\\
&& \hspace*{-1mm}\mbox{C4: }
N_{T_{i}}\in\{N_{\min},N_{\min}\hspace*{-0.5mm}+\hspace*{-0.5mm}1,N_{\min}\hspace*{-0.5mm}+\hspace*{-0.5mm}2,\ldots,
N_{\max}\}, \forall i. \nonumber
\end{eqnarray}
$P_{\max}$ in C1 is the maximum transmit power allowance which puts a
limit on the transmit spectrum mask to control the amount of out-of-cell
interference in the downlink. C2 is imposed to guarantee that the total
power consumption of the system is less than the maximum power
supply from the power grid, $P_{PG}$, cf. Figure \ref{fig:BS}.
 C3 are
the non-negative constraints on the power allocation variables.
C4 are the combinatorial constraints on the number of activated antennas.

Note that the above optimization problem formulation can be extended
to the case of energy efficiency maximization with imperfect CSI and multiple
users, as shown in the related journal paper
\cite{JR:TWC_large_antennas}.

\section{Solution of the Optimization Problem} \label{sect:solution}

The optimization problem in (\ref{eqn:cross-layer})
is a mixed combinatorial and non-convex optimization problem. To
obtain an optimal solution, an exhaustive search is needed which is
computational infeasible for $ N_{T_{i}}, n_F\gg 1$. To strike a balance between
system performance and solution tractability, we transform the problem
transformation in the next section.

\subsection{Problem Transformation}

We assume that the BS chooses the beamforming vector $\mathbf{
f}_{i}$ to be the eigenvector
corresponding to the maximum eigenvalue of $\mathbf{
h}_{i}\mathbf{ h}_{i}^\dag$, i.e., $\mathbf{
f}_{i}=\frac{\mathbf{h}_{i}}{\norm{\mathbf{ h}_{i}}}$.
 The adopted
beamforming scheme is known as maximum ratio transmission (MRT). Then,
we introduce the following proposition by exploiting the properties of
large numbers of antennas.
\begin{proposition}
The asymptotic channel capacity between the BS and the user on subcarrier
$i$ for $N_{T_{i}}\rightarrow \infty$ with MRT can be approximated by
\begin{eqnarray}\label{eqn:asymptotic channel capacity}
\hspace*{-5mm}C_{i}&\stackrel{(a)}{\approx}&
W\log_2\Big(1+\Gamma_{i}\Big)\stackrel{(b)}{\approx}W\log_2\Big(\Gamma_{i}\Big)\,\,\forall
i\\
\hspace*{-5mm}\mbox{where}\,\,
\Gamma_{i}&=&\frac{P_{i}l g N_{T_{i}}}{WN_0}.
\end{eqnarray}
 \end{proposition}
$(a)$ and $(b)$ are due to the law of large numbers and
$N_{T_{i}}\rightarrow \infty$, i.e., $\lim_{N_{T_{i}}\rightarrow
\infty}\frac{\mathbf{ h}_{i}\mathbf{
h}_{i}^\dag}{N_{T_{i}}}=1$ and $\log_2(1+x)\approx \log_2(x)$
for $x\gg 1$, respectively\footnote{In practice, the value of
$N_{\min}$ is selected such that the law of large numbers holds.}.

It can be observed from (\ref{eqn:asymptotic channel capacity}) that all
the subcarriers have the same asymptotic channel gain due to channel
hardening \cite{book:david_wirelss_com}. Then, we have the following
corollary for the considered resource allocation algorithm design.

\begin{Cor}\label{cor:1}
In the limiting case of large numbers of antennas with MRT, the resource
allocation algorithm is a
chunk-based allocation policy. Specifically,
\begin{eqnarray}
N_{T_i}^* &=&N_T,\,\,\forall i,\\
P_{i}^* &=& P,\,\,\,\,\forall i,
\end{eqnarray}
 \end{Cor}
 where ${P}_i^{*}$ and $N_{T_i}^*$ denote the optimal power allocation
and antenna activation solutions in subcarrier $i$, respectively.
In other words, the BS has no incentive to allocate more resources
to a particular subcarrier, since all the subcarriers have the same
channel gain for $N_{T_i}\rightarrow \infty$.

On the other hand, we need to handle the combinatorial constraint on the
antenna activation. We relax constraint C4 such that $N_{T_i}$ can be a
real value between $N_{\min}$ and $N_{\max}$ instead of an integer
value, i.e., $N_{\min}\le N_{T_i} \le N_{\max},\,\forall i$. This
 yields the following relaxed problem:

\begin{eqnarray}
\label{eqn:cross-layer-transformed}&& \qquad \max_{{\cal P},{\cal A}
}\,\, \sum_{i=1}^{n_F}W \log_2\Big(\frac{P_{i}l g
N_{T_{i}}}{WN_0}\Big) \\
\notag && \hspace*{-6mm}\mbox{s.t.}\,\,\mbox{C1:}
\sum_{i=1}^{n_F}P_{i}\le P_{{\max}}, \quad\mbox{C2: }
U_{TP}({\cal P},{\cal A})\le P_{PG}, \\
\notag &&\hspace*{-1mm}\mbox{C3: }P_{i}\ge 0,\, \forall
i,k,\hspace*{6mm}\mbox{C4: }
N_{{\min}}\le N_{T_i}\le
N_{\max}, \forall i. \nonumber
\end{eqnarray}

Now, we focus on the resource allocation algorithm design for the
relaxed problem. Note that the system performance of the relaxed
problem will serve as an upper bound of (\ref{eqn:cross-layer}). It can
be shown that the relaxed optimization
problem is jointly
concave with respect to (w.r.t.) $ P_{i}$
and $ N_{T_{i}}$, cf. Appendix. Besides, it can be shown that the relaxed
problem satisfies Slater's constraint qualification. As a result, for the relaxed problem,
solving the dual problem
is equivalent to solving the primal  which is
known as strong duality \cite{book:convex}.

\subsection{Dual Problem Formulation}
In this subsection, we solve the power allocation and antenna allocation
optimization problem by solving its dual.
 For this purpose, we first need
the Lagrangian function of the primal problem. Upon rearranging
terms, the Lagrangian can be written as
\begin{eqnarray}&&\notag{\cal
L}(\lambda, \beta,{\cal P},{\cal A})\\ &&\hspace*{-6mm}
=\sum_{i=1}^{n_F}
W\log_2\Big(\frac{P_{i}l g N_{T_{i}}}{WN_0}\Big) -\lambda\Big(
\sum_{i=1}^{n_F}P_{i}- P_{{\max}}\Big)\notag\\
&&\hspace*{-6mm}-\beta\Big(N_{T_{i}}\times
P_{A_C}+\sum_{i=1}^{n_F}\varepsilon P_{i} +P_0 -P_{PG}\Big).
\label{eqn:Lagrangian}
\end{eqnarray}
We note that $\underset{i}{\max}\{N_{T_i}\}$ has been replaced by
$N_{T_i}$ in (\ref{eqn:Lagrangian}) due to Corollary \ref{cor:1}.
$\lambda\ge0$ is the Lagrange multiplier
corresponding to the
maximum transmit power allowance C1.
$\beta$ is the Lagrange multiplier accounting for the power usage from
the power grid. On the other hand, boundary constraints C3
and C4 will be absorbed into the Karush-Kuhn-Tucker (KKT) conditions
when deriving the optimal resource allocation policies in the
following. Thus, the dual problem
 is given by
\begin{eqnarray}
\underset{\lambda,\beta\ge 0}{\min}\ \underset{{\cal
P,A}}{\max}\quad{\cal L}(\lambda,\beta,{\cal
P},{\cal A}).\label{eqn:master_problem}
\end{eqnarray}

 \begin{figure}[t]
\centering
\includegraphics[width=3.5 in]{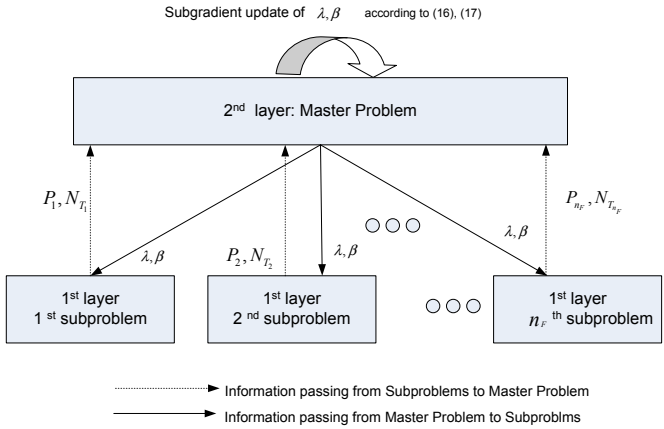}
 \caption{ Illustration of the dual decomposition of a
large-scale problem into a two-layer problem.} \label{fig:layer}
\end{figure}

\subsection{Dual Decomposition }
\label{sect:sub_problem_solution}By Lagrange dual decomposition, the dual
problem is decomposed into two layers: the
first layer consists of $n_F$ subproblems with identical structure; the
second layer is the master problem, cf. Figure \ref{fig:layer}. Then,
the dual problem can be
solved iteratively, where in each iteration the transmitter solves the
 subproblem by using the KKT conditions for a fixed set of Lagrange
multipliers, and the master problem is solved using the gradient method.

\subsubsection{Subproblem Solution}
Using standard optimization
techniques and the KKT conditions, the closed-form resource allocation
policies
   in subcarrier $i$ are obtained
  as:
 \begin{eqnarray}\label{eqn:power1}
\hspace*{-8mm}P_{i}^*&=&P=\Bigg[\frac{{\cal
B}}{\ln(2)(\lambda+\beta \varepsilon)}\Bigg]^+, \quad\forall
i,\\
\label{eqn:antenna_soultion}
\hspace*{-8mm}N_{T_{i}}^*&=& N_T=\Bigg[\frac{{\cal
W}}{\ln(2)P_{A_C}\beta}\Bigg]^{N_{\max}}_{N_{\min}},\forall i.
\end{eqnarray}
The resource allocation solution in (\ref{eqn:power1}) and
(\ref{eqn:antenna_soultion}) can be interpreted as a water-filling
scheme. It can be observed that the BS does not always activate all
the antennas for maximization of the system capacity which is limited
by the total power supply $P_{PG}$ via $\beta $, cf.
(\ref{eqn:antenna_soultion}). Note that since the solution of the
$n_F$ subproblems are identical, the complexity in solving the $n_F$
subproblems can be reduced by a factor of $n_F$ by just focusing on one
subproblem.

\subsubsection{Master Problem Solution}
The dual function is differentiable and, hence, the
gradient method can be used to solve the second layer master problem in
(\ref{eqn:master_problem}) which leads to
\begin{eqnarray}\label{eqn:multipler1}
\hspace*{-3mm}\lambda(m+1)\hspace*{-3mm}&=&\hspace*{-3mm}\Big[\lambda(m)-\xi_1(m)\times
(P_{{\max}}-n_F{P})\Big]^+\hspace*{-3mm},\\
\hspace*{-3mm}\notag\beta(m+1)\hspace*{-3mm}&=&\hspace*{-3mm}\Big[\beta(m)-\xi_2(m)\Big.\\
&&\hspace*{-3mm}\Big.\times
( P_{PG}-N_T\times
P_{A_C}-n_F\varepsilon{P} -P_0)\Big]^+,
\label{eqn:multipler3}
\end{eqnarray}
where index $m\ge 0$ is the iteration index and $\xi_u(m)$,
$u\in\{1,2\}$, are positive step sizes. Since the relaxed
problem is concave in nature, it is
guaranteed that the iteration between master problem and subproblems
converges
to the optimal solution of (\ref{eqn:master_problem}), if the chosen step
 sizes satisfy the infinite travel condition
 \cite{book:convex,Notes:Sub_gradient}.
Then, the updated Lagrange multipliers in
(\ref{eqn:multipler1}), (\ref{eqn:multipler3}) are used for solving
the subproblems in (\ref{eqn:master_problem}) via updating the resource
allocation policies.

Note that the resource allocation solutions obtained in
(\ref{eqn:power1}) and (\ref{eqn:antenna_soultion})
are optimal w.r.t. the relaxed problem in
(\ref{eqn:cross-layer-transformed}) for $N_{T_i}\gg 1$.
\section{Simulation Results}\label{sect:5}
In this section, we study the performance of two algorithms via
simulation. The first algorithm achieves the upper bound performance of
(\ref{eqn:cross-layer}) by solving the relaxed problem in
(\ref{eqn:cross-layer-transformed}) as outlined in Section \ref{sect:cross-Layer_formulation}. For the second algorithm, we
apply a floor function $\lfloor\cdot\rfloor$ to the antenna
allocation solution in (\ref{eqn:antenna_soultion}), i.e.,
$N_{T_{i}}=\lfloor N_{T_{i}}^*\rfloor
$, which achieves a suboptimal performance w.r.t. the original problem
formulation in (\ref{eqn:cross-layer}). For the system setting, there
are $n_F=128$ subcarriers with carrier center frequency $2.5$
GHz and a total system bandwidth of ${\cal B}=5$ MHz. We assume a noise
power of $N_0W=-118$ dBm in each subcarrier.
The desired user is located at a distance of $d_0=500$ m from the BS. Log-normal shadowing with a standard deviation of 8 dB is assumed.
 The multipath fading coefficients of the BS-to-user link are
modeled as i.i.d. Rayleigh random variables with unit
variances. The average system capacity is obtained by assuming capacity-achieving codes and
counting the amount of data which is successfully decoded by the
user averaged over
both shadowing and multipath fading. We assume a static circuit
power consumption of $P_0=$ 40 dBm \cite{CN:static_power}. The
additional power dissipation incurred by each extra antenna for
transmission is $P_{A_C}=30$ dBm \cite{CN:power_consumption_elements}.
On the other hand,
we assume a power efficiency of $40\%$ in the RF PA,
i.e., $\varepsilon=\frac{1}{0.4}= 2.5
$. The maximum and minimum numbers
of active antennas are set to $N_{\max}=500$ and $N_{\min}=10$,
respectively.

 \begin{figure}[t]
\centering
\includegraphics[width=3.5 in]{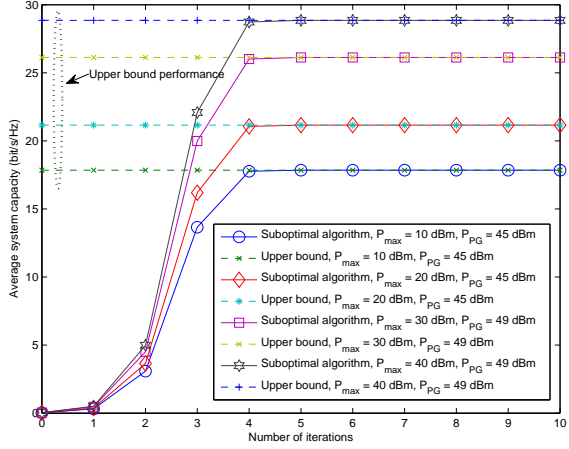}
 \caption{ Average system capacity (bit/s/Hz) versus number of
iterations with different
maximum powers supplied by the power grid, $P_{PG}$, and different
values of maximum transmit power allowance, $P_{\max}$. The dashed
lines
represent an upper bound on maximum average system capacity
for different
cases. }
\label{fig:convergence}
\end{figure}


\subsection{Convergence of Iterative Algorithm }
Figure \ref{fig:convergence} illustrates the evolution of the
proposed iterative algorithm for different values of maximum transmit
power allowance, $P_{\max}$, and maximum power supply, $P_{PG}$.
The results in Figure \ref{fig:convergence} were
averaged over 100000 independent adaptation processes where each
adaptation process involves different realizations of the shadowing and
the multipath fading. It can be observed that
the iterative algorithm converges to values close to the upper bound
performance within 4
iterations for all considered cases. In other
words, a close-to-optimal system capacity can be achieved within a
few iterations on average.

In the following case studies, we focus on the suboptimal algorithm and
set the number of iterations in the algorithm to 5.

\subsection{Average Capacity versus Maximum Power Supply }

\begin{figure}[t]
\centering
\includegraphics[width=3.5 in]{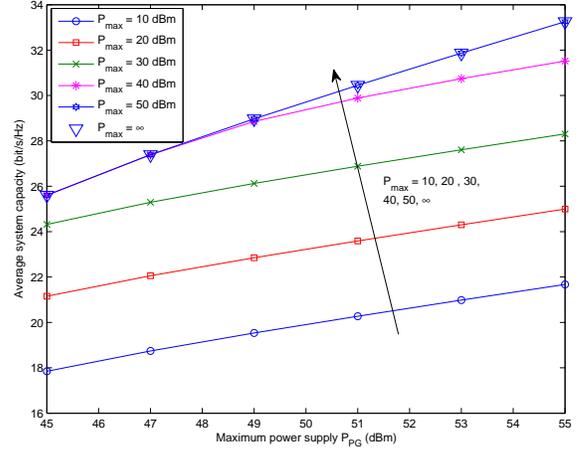}
 \caption{ Average system capacity (bit/s/Hz) versus the
maximum power supplied by the power grid, $P_{PG}$, for different values
of maximum transmit power allowance, $P_{\max}$. }
\label{fig:TP}
\end{figure}

Figure \ref{fig:TP} illustrates the average system capacity (bit/s/Hz)
versus the
maximum power supply from the power grid, $P_{PG}$, for the proposed
suboptimal algorithm. Different values of maximum transmit power
allowance, $P_{\max}$, are considered. It can be observed
that the average system capacity increases for increasing amount of
power supplied by the power grid, since more power is available at the
BS for resource allocation. On the other hand, the system performance also benefits from increasing values of $P_{\max}$. Indeed, from an
optimization theory point of view, a larger value of $P_{\max}$ spans a
larger feasible solution set which results in a possibly higher
objective value. However, there is a diminishing return in performance
as $P_{\max}$ increases, especially in the low power supply regime,
i.e., $P_{PG}<49$ dBm. This is because the system performance is limited
by the
small amount of power $P_{PG}$ supplied by the power grid, instead of
$P_{\max}$.

\begin{figure}[t]
  \centering
  \includegraphics[width=3.5 in]{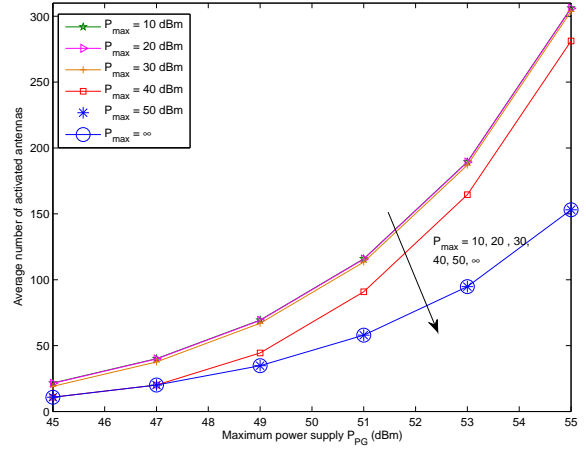}
  \caption{Average number of activated antennas versus the
maximum power supplied by the power grid, $P_{PG}$, for different values
of maximum transmit power allowance,
$P_{\max}$.}\label{fig:power_consumption}
\end{figure}
\subsection{Average Number of
Activated Antennas and Power Consumption versus Maximum Power Supply }

Figure \ref{fig:power_consumption} depicts the average number of
activated antennas versus maximum power supply, $P_{PG}$, for the
proposed suboptimal algorithm. The number of activated antennas
increases rapidly w.r.t. the an increasing $P_{PG}$ due to a larger
amount of available power at the BS. However, the curves in Figure
\ref{fig:power_consumption} have different slopes for different values
of $P_{\max}$. In particular,
the number of activated antennas increases with $P_{PG}$ much faster
for a small value of $P_{\max}$ than for a large value of $P_{\max}$.
 This is because for a small value of $P_{\max}$, the system performance
is always limited by the amount of power radiated in the RF. As a
result, the BS tends to activate more antennas for increasing the
antenna array gain. On the other hand, when the value of $P_{\max}$
increases, the BS will allocate more power to the RF and decrease the
number of activated antennas.
Counterintuitively, the BS does not activate all antennas for maximizing
the average system capacity, when both the PAs and the antennas are
consuming powers from the same power source.

\begin{figure}[t]
  \centering
  \includegraphics[width=3.5 in]{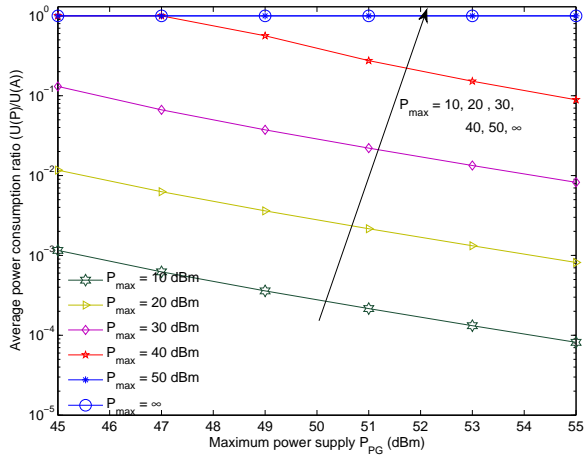}
  \caption{Average power consumption ratio between PAs and antennas
circuits, i.e., $\frac{{\cal U}({\cal P})}{{\cal U}({\cal A})}$,
versus the
maximum power supplied by the power grid, $P_{PG}$, for different values
of maximum transmit power allowance, $P_{\max}$.}\label{fig:ratio}
\end{figure}

Figure \ref{fig:ratio} depicts the power consumption ratio between the
PAs and antennas circuits, i.e., $\frac{{\cal U}({\cal P})}{{\cal
U}({\cal A})}$, versus maximum power supply, $P_{PG}$, for the
proposed suboptimal algorithm. We can see that for a small value of
$P_{\max}$, the BS spends more power on activating antennas when
$P_{PG}$ increases. This is because more antennas will be activated to
enhance the system capacity when the radiated power in the RF reaches
$P_{\max}$. This observation coincides with the results in Figure
\ref{fig:power_consumption}. On the other hand, when $P_{\max}$
increases, the power consumption of both the PAs and the antenna
activation approach the same value for maximizing the system capacity.
Indeed, the BS does not have a higher preference for spending more power
in the PAs than for activating more antennas, or vice versa, when the
constraints on both $P_{\max}$ and $N_{\max}$ in
(\ref{eqn:cross-layer}) become less stringent. On the contrary, the BS
tries to maintain a balance between the power consumption in the PAs and
the antenna circuits. As a result, activating all available antennas may not
be a good solution for system capacity maximization, since the power
consumptions in the antennas and PAs should be balanced.

\section{Conclusion }\label{sect:conclusion}
In this paper, we formulated the resource allocation algorithm
design for MIMO-OFDM networks with large numbers of BS antennas as a
non-convex optimization problem, where a joint power consumption model
for the antennas, power amplifiers, and signal processing circuit unit
was taken into consideration.
An efficient iterative resource allocation algorithm with optimized
power allocation and antenna allocation
policies was proposed. Our simulation results did not only show that the
proposed algorithm converges within a small number of iterations,
but also
 unveiled the trade-off between maximum system capacity and the number
of activated antennas.

\section*{Appendix - Proof of  Concavity of Relaxed Problem}
 Let $\mathbf{H}(C_i)$ and
$\lambda_1, \lambda_2$ be the Hessian matrix of function
$C_i$ and the eigenvalues of $\mathbf{H}(C_i)$, respectively. The
Hessian matrix of function $C_i$ can be expressed as
\begin{eqnarray}\mathbf{H}(C_i)&=&
\begin{bmatrix}
  \frac{-W}{P_i^2\ln(2)} & 0 \\
  0& \frac{-W}{N_{T_i}^2\ln(2)}
\end{bmatrix}.
\end{eqnarray}
Therefore, the corresponding eigenvalues of the
Hessian matrix are given by
\begin{eqnarray} &&\label{eqn:eigvalues}
\lambda_1= \frac{-W}{P_i^2\ln(2)}\quad\mbox{and}\quad
\lambda_2=\frac{-W}{N_{T_i}^2\ln(2)}.
\end{eqnarray}
 Since $\lambda_r\le 0, \,\,
r=\{1,2\}$, so $\mathbf{H}(C_i)$ is a negative semi-definite matrix
and $C_i$ is
jointly concave w.r.t. $P_i$ and $N_{T_i}$. The summation of $C_i$ over
$i$
 preserves the concavity of the objective function in
 \cite{book:convex}. On the other hand, constraints C1-C5 in (\ref{eqn:cross-layer-transformed}) with relaxed
$N_{T_i}$ span a convex feasible set
and thus the transformed problem is a concave optimization problem.

\bibliographystyle{IEEEtran}
\bibliography{OFDMA-AF}

\begin{thebibliography}{10}
\providecommand{\url}[1]{#1}
\csname url@samestyle\endcsname
\providecommand{\newblock}{\relax}
\providecommand{\bibinfo}[2]{#2}
\providecommand{\BIBentrySTDinterwordspacing}{\spaceskip=0pt\relax}
\providecommand{\BIBentryALTinterwordstretchfactor}{4}
\providecommand{\BIBentryALTinterwordspacing}{\spaceskip=\fontdimen2\font plus
\BIBentryALTinterwordstretchfactor\fontdimen3\font minus
  \fontdimen4\font\relax}
\providecommand{\BIBforeignlanguage}[2]{{%
\expandafter\ifx\csname l@#1\endcsname\relax
\typeout{** WARNING: IEEEtran.bst: No hyphenation pattern has been}%
\typeout{** loaded for the language `#1'. Using the pattern for}%
\typeout{** the default language instead.}%
\else
\language=\csname l@#1\endcsname
\fi
#2}}
\providecommand{\BIBdecl}{\relax}
\BIBdecl

\bibitem{book:david_wirelss_com}
D.~Tse and P.~Viswanath, \emph{{Fundamentals of Wireless Communication}},
  1st~ed.\hskip 1em plus 0.5em minus 0.4em\relax {Cambridge University Pres},
  2005.

\bibitem{book:Goldsmith}
A.~Goldsmith, \emph{Wireless Communications}.\hskip 1em plus 0.5em minus
  0.4em\relax Cambridge University Press, 2005.

\bibitem{JR:large_number_antennas}
T.~Marzetta, ``{Noncooperative Cellular Wireless with Unlimited Numbers of Base
  Station Antennas},'' \emph{IEEE Trans. Wireless Commun.}, vol.~9, pp.
  3590--3600, Nov. 2010.

\bibitem{JR:large_antennas-constant-en}
S.~K. Mohammed and E.~G. Larsson, ``{Single-User Beamforming in Large-Scale
  MISO Systems with Per-Antenna Constant-Envelope Constraints: The Doughnut
  Channel},'' \emph{IEEE Trans. Wireless Commun.}, vol.~PP, pp. 1--14, 2012.

\bibitem{JR:LA-detection}
K.~Vishnu~Vardhan, S.~Mohammed, A.~Chockalingam, and B.~Sundar~Rajan, ``{A
  Low-Complexity Detector for Large MIMO Systems and Multicarrier CDMA
  Systems},'' \emph{IEEE J. Select. Areas Commun.}, vol.~26, pp. 473 --485,
  Apr. 2008.

\bibitem{JR:LA4}
P.~Li and R.~Murch, ``{Multiple Output Selection-LAS Algorithm in Large MIMO
  Systems},'' \emph{IEEE Commun. Lett.}, vol.~14, pp. 399--401, May 2010.

\bibitem{CN:LA_measure1}
X.~Gao, O.~Edfors, F.~Rusek, and F.~Tufvesson, ``{Linear Pre-Coding Performance
  in Measured Very-Large MIMO Channels},'' in \emph{Proc. IEEE Veh. Techn.
  Conf.}, Sep. 2011, pp. 1 --5.

\bibitem{CN:LA_measure2}
S.~Payami and F.~Tufvesson, ``Channel measurements and analysis for very large
  array systems at 2.6 ghz,'' in \emph{2012 6th European Conf. on Antennas and
  Propag. (EUCAP)}, Mar. 2012, pp. 433 --437.

\bibitem{CN:power_consumption_elements}
R.~Kumar and J.~Gurugubelli, ``{How Green the LTE Technology Can be?}'' in
  \emph{Intern. Conf. on Wireless Commun., Veh. Techn., Inform. Theory and
  Aerosp. Electron. Syst. Techn.}, Mar. 2011.

\bibitem{CN:energy_efficient_2}
R.~Prabhu and B.~Daneshrad, ``{Energy-Efficient Power Loading for a MIMO-SVD
  System and its Performance in Flat Fading},'' in \emph{Proc. IEEE Global
  Telecommun. Conf.}, Dec. 2010, pp. 1--5.

\bibitem{JR:TWC_large_antennas}
D.~Ng, E.~Lo, and R.~Schober, ``{Energy-Efficient Resource Allocation in OFDMA
  Systems with Large Numbers of Base Station Antennas},'' \emph{IEEE Trans.
  Wireless Commun.}, vol.~11, pp. 3292 --3304, Sep. 2012.

\bibitem{book:convex}
S.~Boyd and L.~Vandenberghe, \emph{{Convex Optimization}}.\hskip 1em plus 0.5em
  minus 0.4em\relax {Cambridge University Press}, 2004.

\bibitem{Notes:Sub_gradient}
S.~Boyd, L.~Xiao, and A.~Mutapcic, ``{Subgradient Methods},'' \emph{{Notes for
  EE392o Stanford University Autumn}}, 2003-2004.

\bibitem{CN:static_power}
O.~Arnold, F.~Richter, G.~Fettweis, and O.~Blume, ``{Power Consumption Modeling
  of Different Base Station Types in Heterogeneous Cellular Networks},'' in
  \emph{Proc. Future Network and Mobile Summit}, 2010, pp. 1--8.

\end{thebibliography}


\begin{thebibliography}{10}
\providecommand{\url}[1]{#1}
\csname url@rmstyle\endcsname
\providecommand{\newblock}{\relax}
\providecommand{\bibinfo}[2]{#2}
\providecommand\BIBentrySTDinterwordspacing{\spaceskip=0pt\relax}
\providecommand\BIBentryALTinterwordstretchfactor{4}
\providecommand\BIBentryALTinterwordspacing{\spaceskip=\fontdimen2\font plus
\BIBentryALTinterwordstretchfactor\fontdimen3\font minus
  \fontdimen4\font\relax}
\providecommand\BIBforeignlanguage[2]{{%
\expandafter\ifx\csname l@#1\endcsname\relax
\typeout{** WARNING: IEEEtran.bst: No hyphenation pattern has been}%
\typeout{** loaded for the language `#1'. Using the pattern for}%
\typeout{** the default language instead.}%
\else
\language=\csname l@#1\endcsname
\fi
#2}}

\bibitem{CN:energy_consumption}
G.~P. Fettweis and E.~Zimmermann, ``{ICT Energy Consumption - Trends and
  Challenges},'' in \emph{Proc. IEEE Intern. Conf. on Acoustics, Speech and
  Signal Process.}, Sept. 2008.

\bibitem{Tech:EARTH}
\BIBentryALTinterwordspacing
``{Energy Aware Radio and neTwork tecHnologies (EARTH)}.'' [Online]. Available:
  \url{https://www.ict-earth.eu/default.html,}
\BIBentrySTDinterwordspacing

\bibitem{JR:capacity_maximization}
J.~Jang and K.~B. Lee, ``{Transmit Power Adaptation for Multiuser OFDM
  Systems},'' \emph{IEEE J. Select. Areas Commun.}, vol.~21, pp. 171 -- 178,
  Feb. 2003.

\bibitem{JR:capacity_maximization2}
G.~Song and Y.~Li, ``{Cross-layer Optimization for OFDM Wireless Networks-Part
  I: Theoretical Framework},'' \emph{IEEE Trans. Wireless Commun.}, vol.~4, pp.
  614 -- 624, Mar. 2005.

\bibitem{JR:energy_efficient_2}
G.~Miao, N.~Himayat, and G.~Li, ``{Energy-Efficient Link Adaptation in
  Frequency-Selective Channels},'' \emph{IEEE Trans. Commun.}, vol.~58, pp.
  545--554, Feb. 2010.

\bibitem{CN:EM_1}
C.~Isheden and G.~Fettweis, ``{Energy-Efficient Multi-Carrier Link Adaptation
  with Sum Rate-Dependent Circuit Power},'' in \emph{Proc. IEEE Global
  Telecommun. Conf.}, Dec. 2010, pp. 1 --6.

\bibitem{CN:EM_2}
R.~Prabhu and B.~Daneshrad, ``{An Energy-Efficient Water-Filling Algorithm for
  OFDM Systems},'' in \emph{Proc. IEEE Intern. Commun. conf.}, May 2010, pp. 1
  --5.

\bibitem{JR:Roger_OFDMA}
C.~Y. Wong, R.~S. Cheng, K.~B. Letaief, and R.~D. Murch, ``{Multiuser OFDM with
  Adaptive Subcarrier, Bit, and Power Allocation},'' \emph{IEEE J. Select.
  Areas Commun.}, vol.~17, pp. 1747--1758, Oct 1999.

\bibitem{CN:static_power}
O.~Arnold, F.~Richter, G.~Fettweis, and O.~Blume, ``{Power Consumption Modeling
  of Different Base Station Types in Heterogeneous Cellular Networks},'' in
  \emph{Proc. Future Network and Mobile Summit}, 2010, pp. 1--8.

\bibitem{JR:CM_PM_dual}
N.~Papandreou and T.~Antonakopoulos, ``{Bit and Power Allocation in Constrained
  Multicarrier Systems: The Single-User Case},'' \emph{{EURASIP J. Advances
  Signal Process.}}, vol. 2008, 2008, article ID 43081.

\end{thebibliography}
\end{document}